\documentstyle[aps,graphicx,prl]{revtex}

\newcommand{\myscalebox}[1]{\scalebox{0.43}[0.43]{#1}}
   
\begin{document}

%%%%%%%%%enable the following command for two-column mode%%%%%%%%%%%%%%%%%%%%%
\twocolumn[\hsize\textwidth\columnwidth\hsize\csname@twocolumnfalse\endcsname

\title{The number of guards needed by a museum: a phase transition in
  vertex covering of random graphs}
\author{Martin Weigt and Alexander K. Hartmann}
\address{Institute for Theoretical Physics,
      University of G\"ottingen, Bunsenstr. 9, 37073 G\"ottingen, Germany\\
      E-mail: \texttt{weigt/hartmann@theorie.physik.uni-goettingen.de}
        }

\date{\today}
\maketitle
 
\begin{abstract}
In this letter we study the NP-complete vertex cover problem on finite
connectivity random graphs. When the allowed size of the cover set is
decreased, a discontinuous transition in solvability and typical-case
complexity occurs. This transition is characterized by means of exact
numerical simulations as well as by analytical replica calculations.
The replica symmetric phase diagram is in excellent agreement with
numerical findings up to average connectivity $e$, where replica
symmetry becomes locally unstable. \\
{\bf Keywords (PACS-codes)}: 
General studies of phase transitions (64.60.-i),
Classical statistical mechanics (05.20.-y),
Combinatorics (02.10.Eb)
\pacs{64.60.-i, 05.20.-y, 02.60.Ph, 02.10.Eb}
\end{abstract}
\vskip.5pc]

\narrowtext
Imagine you are director of an open-air museum situated in a large
park with numerous paths. You want to put guards on crossroads 
to observe every path, but in order to economize
costs you have to use as few guards as possible. Let $N$ be the number
of crossroads, $X<N$ the number of guards you are
able to pay.  Then there are $N\choose X$ possibilities of
putting the guards, but the most ``configurations'' will lead to
unobserved paths. Deciding whether there exists any perfect solution
or finding one can thus take a time growing
exponentially with $N$. In fact, this problem is one of the six basic
NP-complete problems \cite{GaJo}, namely {\em vertex cover} (VC).  
It is widely believed that no algorithm can be found which
solves our problem substantially faster than exhaustive search
for any configuration of the paths. 

Similar combinatorial decision problems have been found to show
interesting phase transition phenomena. These occur in their
solvability and, even more surprisingly, in their typical-case
algorithmic complexity, {\it i.e.} the dependence of the median
solution time on the system size \cite{review}. {\it E.g.} in
satisfiability (SAT) problems a number of Boolean variables has to
simultaneously satisfy many logical clauses. When the number of these
(randomly chosen) clauses exceeds a certain threshold, the solvability
of the full problem undergoes a sharp transition from almost always
satisfiable to almost always unsatisfiable \cite{MiSeLe}.  The
instances which are hardest to solve are found in the vicinity of the
transition point.  Far away from this point the solution time is much
smaller, as a formula is either easily fulfilled or hopelessly
over-constrained. The typical solution times in the under-constrained
phase are even found to depend only polynomially on the system size.
Recently, insight coming from a statistical mechanics perspective on
these problems \cite{MoZe} has lead to a fruitful cooperation with
computer scientists, and has shed some light on the nature of this
transition \cite{nature}. Frequently, the methods of statistical
mechanics allow to obtain more insight than the classical tools of
computer science or discrete mathematics.

This is also true for the above mentioned VC problem. After having
introduced the VC model and reviewed some previously known rigorous
results, we present numerical evidence for the existence of a phase
transition in its solvability which is connected to an exponential
peak in the typical case complexity. Due to the much simpler
geometrical structure, many features of this transition can be
understood much more intuitively than for SAT. In addition, we will
see that the replica-symmetric \cite{replica-symmetry} theory
correctly describes the phase transition up to an average connectivity
$e$. This is a fundamental difference to previously studied models
with discontinuous transitions; see \cite{BiMoWe} for the example of
3-satisfiability where replica symmetry breaking is necessary to
calculate the transition threshold.

Let us reformulate our problem in a mathematical way: Take any graph
$G=(V,E)$ with $N$ vertices $i\in V=\{1,2,...,N\}$ (the crossroads in
the above example) and edges $(i,j)\in E\subset V\times V$ (the
paths). We consider undirected graphs, so with $(i,j)\in E$ we also
have $(j,i)\in E$. A vertex cover is a subset $V_{vc}\subset V$ of
vertices such that for all edges $(i,j)\in E$ there is at least one of
its endpoints $i$ or $j$ in $V_{vc}$ (the path is observed). 
We call the vertices in $V_{vc}$ covered, whereas the
vertices in its complement $V\setminus V_{vc}$ are called uncovered.
Please note that the VC of a disconnected graph is consequently given 
by the union of the VCs of its connected components. 

Also {\em partial VCs} $U\subset V$ are considered, where there are
some uncovered edges $(i,j)$ with $i\notin U$ and $j \notin U$. The
task of finding the minimum number of uncovered edges for given graph
$G$ and the cardinality $|U| \equiv X$ is an {\it optimization
problem}. We have already mentioned that the corresponding 
{\em decision problem} if a VC of fixed cardinality $X$ does exist or 
not, belongs to the basic NP-complete problems.

In order to be able to speak of typical or average cases, we have to
introduce some probability distribution over graphs. We investigate
random graphs $G_{N,c/N}$ with $N$ vertices and edges $(i,j)$ (with
$i\neq j$) which are drawn randomly and independently with probability
$c/N$. Thus the expected edge number equals ${N\choose 2}c/N =
cN/2+O(1)$. The average connectivity $c$ remains finite in the limit
$N\to\infty$ of infinitely large graphs.  For $c<1$, these graphs are
known to be decomposed into $O(N)$ connected components of typical
size $O(1)$, whereas for $c>1$ a giant component appears which unifies
$O(N)$ vertices \cite{ErRe}. For a more recent and complete 
introduction see \cite{Bo}.

As an element of a VC $V_{vc}$ typically covers
$O(c)$ edges, the minimal cover size $X_{min}$ is also expected to 
be of $O(N)$, $X_{min}=x_{N,c}N$. In fact, there are rigorous lower 
and upper bounds on $x_{N,c}$ which are valid for almost all random
graphs. To our knowledge, the best bounds are given in \cite{Ga,Ha},
see Fig. \ref{figXBC} for a comparison with our results.
The exact asymptotics for large connectivities $c\gg 1$ is also known, 
Frieze proved that \cite{Fr}
\begin{equation}
  \label{eq:asympt}
  x_{N,c} = 1- \frac{2}{c} (\log c -\log\log c -\log 2+1)
        +o(\frac{1}{c})
\end{equation}
for almost all graphs $G_{N,c/N}, \ N\to\infty$.
 
It is however not clear, if a sharp threshold
$x_c(c)=\lim_{N\to\infty} \overline{x_{N,c}}$ does exist at finite
$c$, with the over-bar denoting the average over the random graph
ensemble at fixed $N$ and $c$.  In order to get some intuition on this
point we have started our work with exact numerical simulations. 
Analytic results are presented below.

Using an exact branch-and-bound algorithm \cite{Lawler66,Tarjan77} all
optimal configurations at fixed $X$ are enumerated: As each vertex is
either covered or uncovered, there are ${N\choose X}$ possible
configurations which can be arranged as leaves of a binary
(backtracking) tree. At each node of the tree, the two subtrees
represent the subproblems where the corresponding vertex is either
covered or uncovered. A subtree will be omitted if its leaves can be
proven to contain less covered edges than the best of all previously
considered configurations.  The order of the vertices within the
levels of the tree is given by their current connectivity, {\it i.e.}
only neighbors are counted which are not yet included into the cover
set. Thus, the first descent into the tree is equivalent to the greedy
heuristic which iteratively covers vertices by always taking the
vertex with the highest current connectivity.

First results are exposed in Fig. \ref{figPEX}: The probability of 
finding a vertex cover of size $xN$ in a random graph $G_{N,c/N}$
is displayed for $c=2$ and several values of $N$, analogous
results have been obtained for other values of $c$. The drop of the
probability from one for large cover sizes to zero for small cover sets
obviously sharpens with $N$, so that a jump at a well-defined
$x_c(c)$ is to be expected in the large-$N$ limit: for $x>x_c(c)$
almost all random graphs with $cN$ edges are coverable with $xN$
vertices, below $x_c(c)$ almost no graphs have such a VC. 
Fig. \ref{figPEX} also shows the minimal fraction $e$ of uncovered edges
as a function of $x$ for the partial covers. 
It vanishes for $x>x_c(c)$, whereas it remains positive for $x<x_c(c)$.

It is also interesting to measure the median computational effort, as
given by the number of visited nodes in the backtracking tree, in
dependence on $x$ and $N$.  The curves, which are given in
Fig. \ref{figTimeX}, show a pronounced peak near the threshold
value. Inside the coverable phase, $x>x_c(c)$, the computational cost
is growing only linearly with $N$, and in many cases the greedy
heuristic is already able to cover all edges by 
covering $xN$ vertices. Below
the threshold, $x < x_c(c)$, the computational effort is clearly
exponential in $N$, but becomes smaller and smaller if we go away from
the threshold.  This easy-hard-easy scenario resembles very much the
typical-case complexity pattern of 3SAT \cite{nature}, and deserves
some analytical investigation.

To achieve this, we use the strong similarity between combinatorial 
optimization problems and statistical mechanics. In the first case, a
cost function depending on many discrete variables has to be
minimized, {\it e.g.} the number of uncovered edges is such a cost
function for vertex cover. This is equivalent to zero temperature
statistical mechanics, where the Gibbs weight is completely
concentrated in the ground states of the Hamiltonian. As the
local variables for VC are binary because a vertex is either
covered or uncovered, we may give a canonical one-to-one mapping 
of the vertex cover problem to an Ising model:
for any subset $U\subset V$ we set $S_i=+1$ if $i\in U$, and $S_i=-1$
if $i\notin U$. The edges are encoded in the adjacency matrix 
$(J_{ij})$:  an entry equals 1 iff $(i,j)\in E$, and $J_{ij}=0$ else. 
$(J_{ij})$ is thus a symmetric random matrix with independently and
identically distributed entries in its lower triangle. The
Hamiltonian, or cost function, of the system counts the number of 
edges which are not covered by the elements of $U$,
\begin{equation}
  \label{hamiltonian}
  H = \sum_{i<j} J_{ij} \delta_{S_i,-1} \delta_{S_j,-1},
\end{equation}
and has to be minimized under the constraint $|U|=xN$, which, in terms
of Ising spins, reads
\begin{equation}
  \label{magn}
  \frac{1}{N} \sum_i S_i = 2x-1.
\end{equation}
The resulting ground state energy $e_{gs}(c,x)$ equals zero iff 
the 
graph is coverable with $xN$ vertices.

We want to skip the details of the calculation, as these go
beyond the scope of this letter. A detailed technical description will
be presented elsewhere \cite{HaWe}. We only mention the main steps:\\
{\it (i)} We introduce a positive formal temperature $T$ and calculate the
  canonical partition function $Z=\sum_{{\cal{C}}_x}\exp\{-H/T\}$
  where the sum is over all configurations $\{S_i\}_{i=1,...,N}$ which
  satisfy (\ref{magn}).\\
{\it (ii)} We are interested in the disorder-averaged free-energy density 
  $f(c,x)=-\lim_{N\to\infty}TN^{-1}\overline{\ln Z}$, which we
  calculate using the replica method, closely following the scheme
  proposed in \cite{Mo}. Within the replica symmetric framework, 
  this free energy self-consistently depends on
  the order parameter $P(m)$ which is the histogram of local 
  magnetizations  $m_i=\langle S_i \rangle$. 
  $\langle\cdot\rangle$ denotes the thermodynamic average.\\
{\it (iii)} The ground states are recovered by sending $T\to 0$.
  In this limit, one has to take care of the scaling of the order 
  parameter with $T$, which is different below and above $x_c(c)$. 
  For a similar reasoning in the case of 3SAT see also
  \cite{BiMoWe}.\\
{\it (iv)} Both equations for $x<x_c(c)$ and $x>x_c(c)$ tend to the same
  limit for $x\to x_c(c)$. At the threshold, the resulting 
  self-consistency equation can be solved analytically.

From this solution, many properties of the threshold VCs can be
read off. The first is of course the value of the threshold itself:
\begin{equation}
  \label{x_c}
  x_c(c) = 1-\frac{2W(c)+W(c)^2}{2c}
\end{equation}
with the Lambert-W-function $W$ \cite{lambertw}.
The result for $x_c(c)$ is displayed in Fig. \ref{figXBC}
along with numerical data obtained by a variant of the
branch-and-bound algorithm. For relatively small connectivities $c$ 
perfect agreement is found. We also have compared (\ref{x_c}) with
rigorous bounds obtained from counting VCs for small connected
components having up to 7 vertices, which are very precise for small
c (e.g. 0.999997N vertices are taken into account for c=0.1). Also
here, perfect coincidence was found. 

For larger $c$ systematic deviations of (\ref{x_c}) from numerical
results occur, it even violates the asymptotic form (\ref{eq:asympt}).
For $c>e$, the replica symmetric solution becomes instable, and we
find a continuous appearance of a replica symmetry broken solution;
work is in progress on this point \cite{HaWe}.  We conjecture, that
the replica symmetric result (\ref{x_c}) is exact for $c\leq e$,
whereas it gives a lower bound for $c>e$ \cite{maxf}. Please note that
this point is situated well beyond $c=1$ where the giant component
appears. Neither analytically nor numerically, we have found any
influence of the giant component on the vertex covers. This is 
significantly different from Ising models on random graphs as studied 
in \cite{KaSo}.

Besides the value of $x_c(c)$, the replica symmetric solution also
contains structural information. One important phenomenon is a partial
freezing of degrees of freedom. For a given random graph, there exists
typically an exponential number of minimal VCs, thus the entropy
density is finite. On the other hand, a fraction $b_+(c)$ of the
vertices will be covered in {\it all} minimal VCs, thus forming a {\em
covered backbone}, other vertices will never be covered and are
collected in the {\em uncovered backbone} which has size $b_-(c)N$:
\begin{eqnarray}
  \label{eq:bb}
  b_-(c) &=& \frac{W(c)}{c}\nonumber\\
  b_+(c) &=& 1-\frac{W(c)+W(c)^2}{c}
\end{eqnarray}
In Fig. \ref{figXBC} the total backbone size $b_c(c)=b_-(c)+b_+(c)$
is compared with numerical data, again very good agreement is found in
the range of validity of replica symmetry.

For small $c$, the uncovered backbone is large, which is mainly due
to isolated vertices which have to be uncovered in minimal VCs. The
simplest structure showing a covered backbone are subgraphs with
three vertices and two edges. In the minimal VC of this subgraph, the
central vertex is covered, thus belonging to the covered backbone,
the other two are uncovered, thus belonging to the uncovered backbone.
The simplest non-backbone structures are components with only two
vertices and one edge, because the vertices have no unique covering 
state.

These backbones appear discontinuously at the threshold because inside
the coverable phase the backbone is empty. The proof is simple
($x>x_c(c)$ fixed):\\ 
{\it (i)} Assume that there is a non-empty uncovered backbone, with $i$
  being an element. Now take any minimal cover $V_0$. It can be
  extended by covering arbitrarily chosen $(x-x_c(c)) N$ vertices out of
  $V\setminus V_0$, {\it e.g.} vertex $i$, which is a contradiction to
  our assumption.\\
{\it (ii)}  Assume now a non-empty covered backbone, with $i$ being an
  element. Then $i$ has to be an element of $V_0$. As the connectivity
  of $i$ is almost surely smaller than or equal to $O(\log N)$,
  all uncovered neighbors of $i$ can be covered by some of the 
  $(x-x_c(c)) N$ covering marks (for $N$ sufficiently large), and $i$
  can be uncovered without uncovering the graph. This is again a
  contradiction to our assumption.

To summarize, we have investigated the vertex cover problem on random
graphs by means of exact numerical simulations and analytical replica
calculations. A sharp transition from a coverable to an uncoverable
phase is found by decreasing the permitted size of the cover set. This
transition coincides with a change of the typical case complexity from
linear to exponential growth in $N$ and the discontinuous appearance
of a frozen-in backbone. The complete RS solution was
given for $c<e$, it is found to be in perfect agreement with
numerical results. For $c>e$ the behavior is less clear as replica
symmetry breaking occurs.

Also the behavior inside the coverable and the uncoverable phases 
is of some interest. There
the use of variational techniques similar to those proposed in
\cite{BiMoWe} could be of great help.

The authors are grateful to J.A. Berg for critically reading the
manuscript. Financial support was provided by the DFG ({\em Deutsche 
Forschungsgemeinschaft}) under grant Zi209/6-1.

\newcommand{\captionPEX}
{Probability $P_{cov}(x)$  that a cover exists for a random graph
 ($c=2$) as a function of
  the fraction $x$ of covered vertices. The result is shown for three
  different system sizes $N=25,50,100$ (averaged over $10^4$ - $10^3$ 
  samples). Lines are guides to the eyes only. In the left part, where
  $P_{cov}=0$, the energy $e$ (see text) is positive. The inset 
  enlarges this result the region $0.3\le x \le 0.5$.}

\newcommand{\captionTimeX}
{Time complexity of vertex cover: 
Median number of nodes visited in the backtracking tree as a
  function of the fraction $x$ of covered vertices for graph sizes
  $N=20,25,30,35,40$ ($c=2$). 
The inset shows the  region below the threshold with logarithmic
scale, including also data for $N=45,50$. The fact that in this 
representation the lines are equidistant
shows that the time complexity grows exponentially with $N$.}

\newcommand{\captionXBC}
{Phase diagram: critical fraction $x_c$ of covered vertices as a
function of the edge density $c$. For $x>x_c(c)$, almost all graphs 
have VCs with $xN$ vertices, while they have almost surely no VC 
with $x<x_c(c)$. The solid line shows our analytic result. 
The rigorous bounds are given by dot-dashed \protect\cite{Ga} 
resp. dashed \protect\cite{Ha} lines. The vertical line is at $c=e$. 
The circles
represent the results of the numerical simulations. Error bars are
much smaller than symbol sizes. All
numerical values were calculated from finite-size scaling fits of
$x_{N,c}$ using functions $x_{N,c}=x_c(c)+aN^{-b}$. 

The inset shows the
threshold backbone size $b_c$ as a function of $c$: analytical results
are given by the solid line, numerical data by the error bars.}

\begin{figure}[htb]
\begin{center}
\myscalebox{\includegraphics{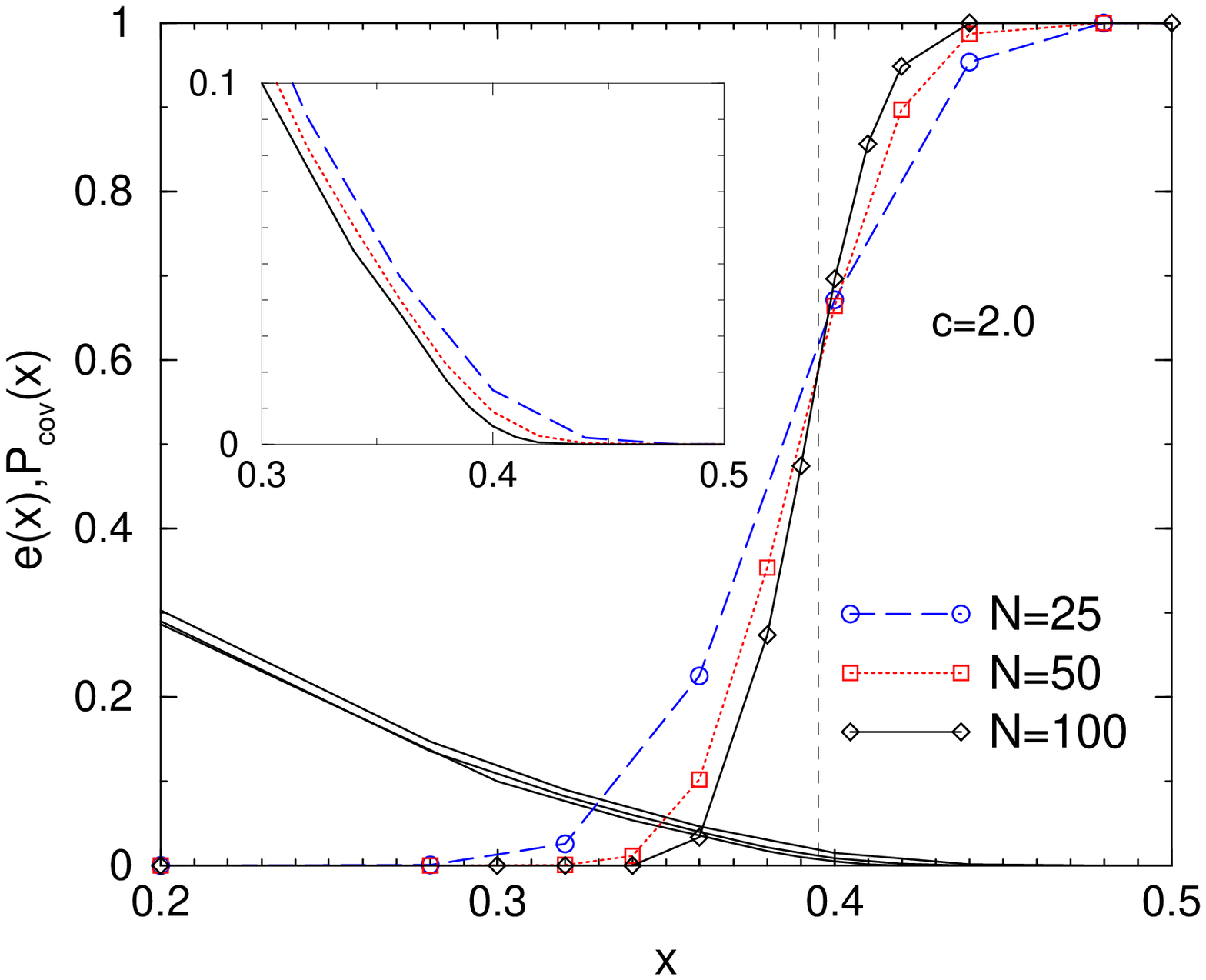}}
\end{center}
\caption{\captionPEX}
\label{figPEX}
\end{figure}

\begin{figure}[htb]
\begin{center}
\myscalebox{\includegraphics{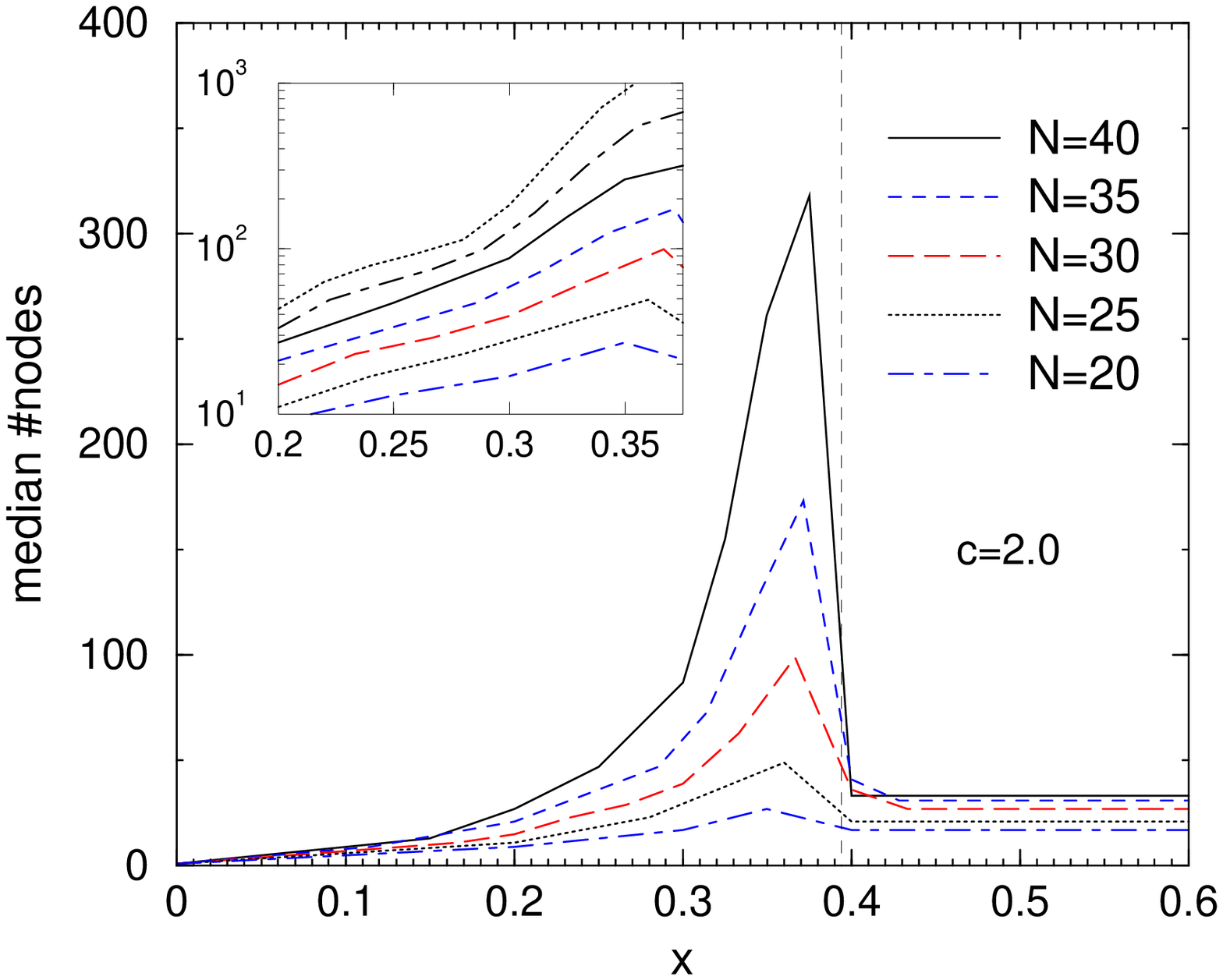}}
\end{center}
\caption{\captionTimeX}
\label{figTimeX}
\end{figure}

\begin{figure}[htb]
\begin{center}
\myscalebox{\includegraphics{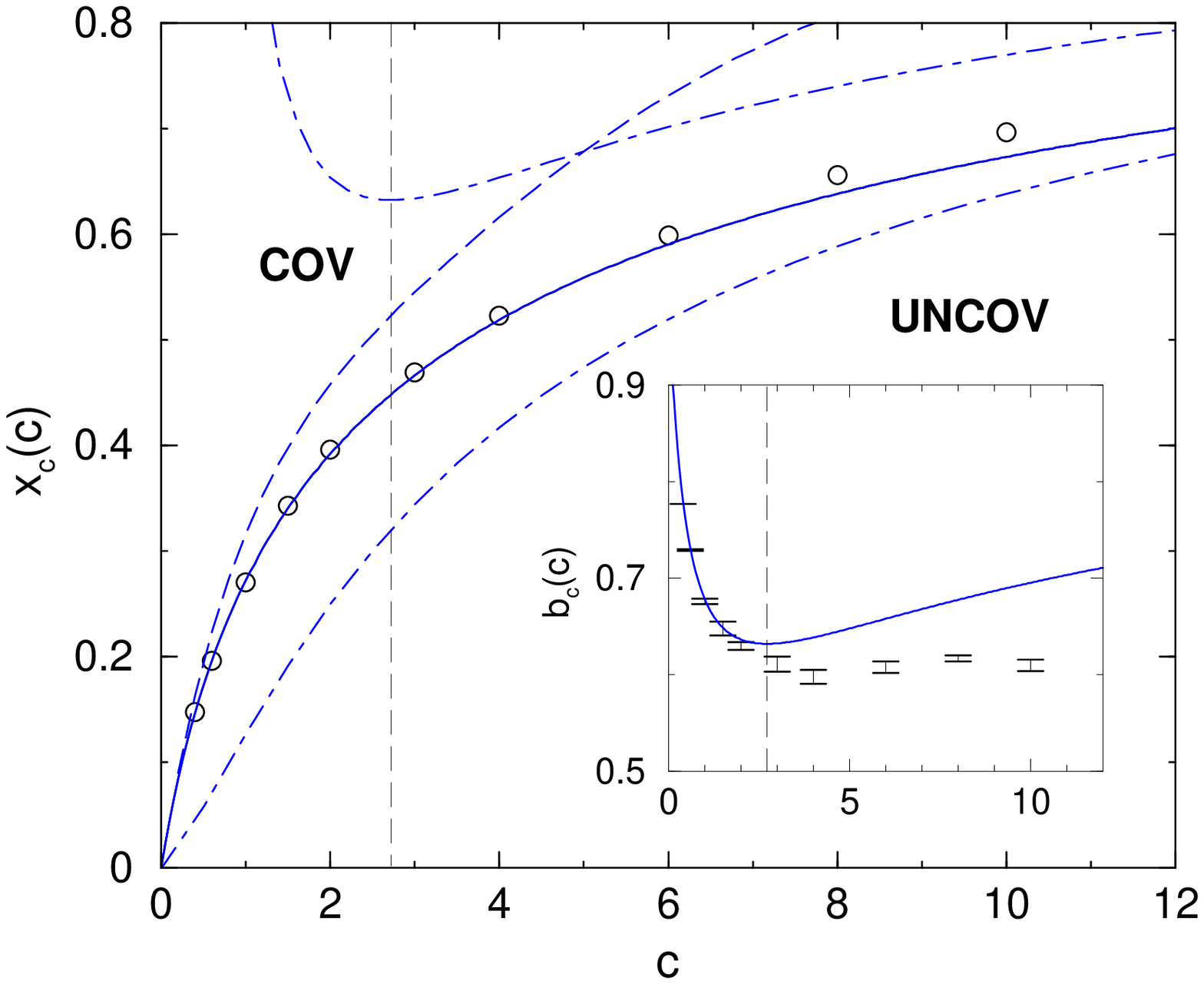}}
\end{center}
\caption{\captionXBC}
\label{figXBC}
\end{figure}


\begin{thebibliography}{99}
\bibitem{replica-symmetry} Informal: replica symmetry assumes that the
  VCs are collected in one cluster in configuration space, whereas
  replica-symmetry breaking corresponds to a large number of clusters.
\bibitem{maxf} Imagine there are two different values for $x_c$
  resulting from different solutions of the saddle point equation.
  In between both, one would already imply a positive ground state
  energy, whereas the other would still be zero. Due to the standard
  procedure of taking saddle points with larger free energy in the
  replica approach, the positive energy solution has to be preferred,
  and hence the larger threshold.
\end{thebibliography}
\end{document}